# Comparison of Advance Tree Data Structures

Parth Patel
Student - Master of Engineering
CSED, Thapar University
Patiala, India

Deepak Garg
Associate Professor
CSED, Thapar University
Patiala, India

## ABSTRACT

B-tree and R-tree are two basic index structures; many different variants of them are proposed after them. Different variants are used in specific application for the performance optimization. In this paper different variants of B-tree and R-tree are discussed and compared. Index structures are different in terms of structure, query support, data type support and application. Index structure's structures are discussed first. B-tree and its variants are discussed and them R-tree and its variants are discussed. Some structures example is also shown for the more clear idea. Then comparison is made between all structure with respect to complexity, query type support, data type support and application.

## 1. INTRODUCTION
Index is a data structure enables sub linear time lookup and improves performance of searching. A data store contains N objects we want to retrieve one of them based on value. Number of operation in worst case is $\Omega$ (n). In real life data store contain millions of data for real world objects and searching is most common and always use to retrieval of data. So, to improve this performance indexing of data is required. Many index structure have O(log(N)) complexity and in some application it is possible to achieve (O(1)). There are many different index structures use for this purpose. Main goal of indexing is to optimize the speed of query [20]. For any type of search or retrieval of information we ask a query and query is process by database system or search engine internally process query on database of different content. Different index structures are there. B-tree and R-tree are basic and most common index structures. They have some disadvantage so their variants are introduced and used. Actually data are not only in linear form they are multidimensional and different type like document, media etc. Main goal of indexing is to optimize the speed of query. For any type of search or retrieval of information we ask a query and query is process by database system or search engine internally process query on database of different content. A number of indexing structure are proposed for various application. A good index structure has ability to collect similar data into same portion. Index structure classifies data into the same cluster for consistency. Some of the index structures that are widely used and some are more application or query type specific. In this paper introduction to basic data structure B-tree and R-tree their application, advantage and disadvantage. What are the changes made into the basic index structure for improvement?

The paper is organized as follows in section 2 structure of B-tree and R-tree are described. Section 3 variants of B-tree and R-tree are discussed. Section 4 comparison between different index structures based on their performance and their application. In section 5 conclusion and future scope.

## 2. B-TREE AND R-TREE
The data structure which was proposed by Rudolf Bayer for the Organization and Maintenance of large ordered database was B-tree [12]. B-tree has variable number of child node with predefined range. Each time node is inserted or deleted internal nodes may join and split because range is fix. Each internal node of B-tree contains number of keys. Number of keys chosen between d and 2d, d is order of B-tree. Number of child node of any node is d+1 to 2d+1. B-tree keeps record in sorted order for traversing. The index is adjusted with recursive algorithm. It can handle any no of insertion and deletion. After insertion and deletion it may require rebalancing of tree.

As per Knuth's definition [6], B-tree of order n (maximum number of children for each node) is satisfied following property:

1. Every node has at most n children.
2. Every node has at least n/2 children.
3. The root has at least two children if it is not a child node.
4. All leaf node at the same level.
5. A non-Leaf node have n children contains n-1 keys.

It best case height of B-tree is logmn and worst case height is logm/2n. Searching in B-tree is similar to the binary search tree. Root is starting then search recursively from top to bottom. Within node binary search is typically used. Apple's file system HFS+, Microsoft's NTFS [8] and some Linux file systems, such as btrfs and Ext4, use B-trees. B++tree, B* tree and many other improved variants of B-tree is also proposed for specific application or data types. B-tree is efficient for the point query but not for range query and multi-dimensional data [4].

Spatial data cover space in multidimensional not presented properly by point. One dimensional index structure B-tree do not work well with spatial data because search space is multidimensional. R-tree was proposed in 1982 by Antonin Guttman. It is dynamic index structure for the spatial searching [1]. It represent data object in several dimension. It is height balanced tree like B-tree. Index structure is dynamic; operation like insertion and deletion cam be intermixed with searching.

Let M be the maximum number of entries in one node and minimum number of entries in a node is m$\leq$ M/2. R-tree satisfies following properties [1]:

1. Each leaf node(Unless it is root) have index record between m and M.
2. Each index record (I, tuple- identifier) in a leaf node. I is smallest rectangle represented by the indicated tuple and contains the n- dimensional data object.
3. Each non-leaf (unless it is root) has children between m and M.
4. Each entry in non-leaf node (I, child pointer), I contain the rectangle in the child node is the smallest rectangle.

5. The root node (unless it is children) at least two children.
6. All leaves appear on the same level.

Fig 1 and Fig 2 show structure of R-tree and relation exist between its rectangles [1].

The searching is similar to the B-tree. More than one sub tree under a node may need to be searched, hence not guarantee worst-case performance. Inserting records is similar to insertion in B-tree. New records are added and overflow result into split and splits propagate up the tree. Relational database systems that have conventional access method, R-tree is easy to add. R-tree give best performance when it is 30-40 % full because more complex balancing is require for spatial data. Disadvantage of space wastage in R-tree variant of R-tree were also proposed. R+-tree, R*-tree, Priority R-tree, Hilbert tree, X-tree etc.

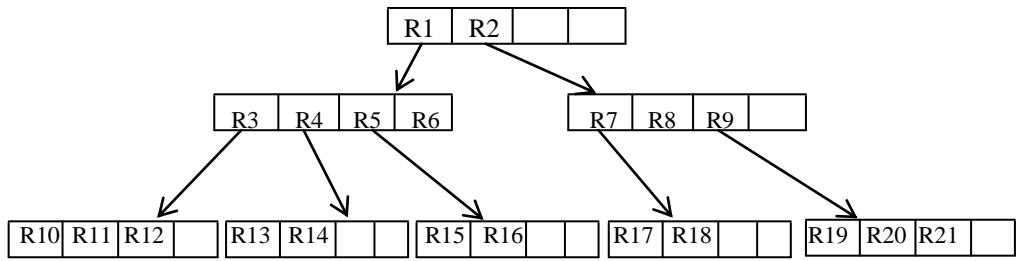

**Fig 1: Structure of R-tree**

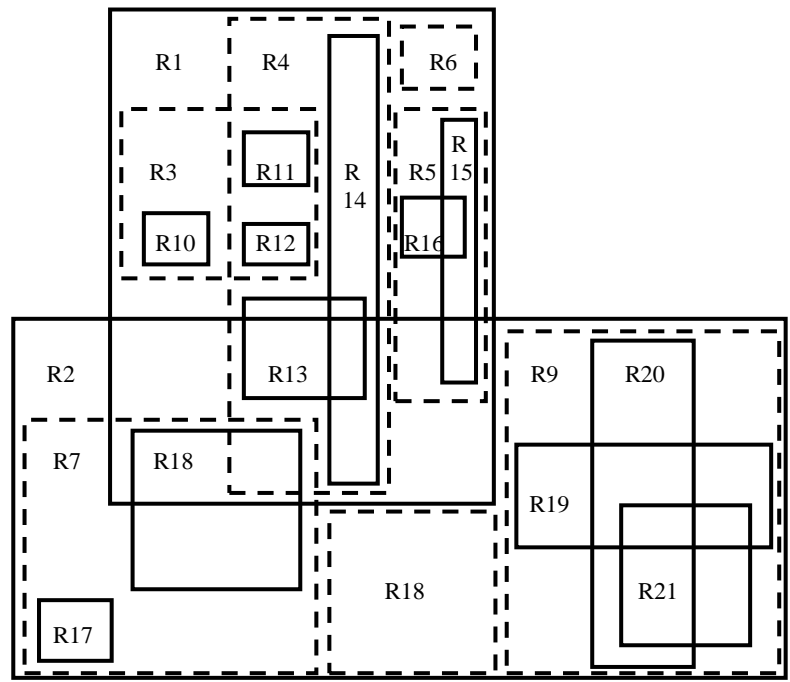

**Fig 2: Overlapping relation between rectangles**

# 3. VARIANTS OF B-TREE AND R-TREE

## 3.1 Variants of B-tree

B+-tree is similar to the B-tree the difference is all records are stored at leaf level and only keys stored in non-leaf nodes. Order of B+-tree b is capacity of node, number of children to a node referred as m, constrained for internal node that ($[b]2]\le m\le b$. The root allowed having as few as 2 children; the numbers of keys are at least b-1 and at most b. No paper on B+-tree but a survey of B-tree also covering B+-tree [6]. Figure 3 shows B+-tree example.

B+-tree is widely use in most of the rational database system for metadata indexing and also useful for the data stored in the RAM.

To keep internal node more densely packed B*-tree balance more internal neighbor nodes [6]. This require non-root node to be at least 2/3 fill. When both nodes are full they split into three, single node gets full then it shares with the next node.

UB-tree [8] is proposed for storing and retrieving the multidimensional data. It is like B+-tree but records are stored according to Z-order or called Morton order. The algorithm provided for the range search in the multidimensional point data is exponential with dimension so not feasible.

H-tree is a special index structure similar to B-tree but use for directory indexing. It has constant depth of one or two levels and do not require balancing, use a hash of a file name. It is use in Linux file system ext3 and ext4.

ST2B-tree: A Self-Tunable Spatio-Temporal B+-tree Index for Moving Objects [9]. It is built on B+-tree without change in insertion and deletion algorithms. It index moving objects as 1d data points. 1d key has two components: KEYtime and KEYspace. Object is updated once in a time ST2B-tree splits tree into two sub trees. Logically it splits B+-tree and each sub tree assign a range. A moving object is a spatial temporal point in natural space. For index in the space data space is partitioned into the disjoint regions in terms of reference point's distance. In this structure reference point and grid granularity are tunable. ST2B-tree meets two requirements:

1. Discriminate between regions of different densities.
2. Adapt to density and distribution changes with time.

Use B+-tree for the multinational data need to reduce dimension and data density and granularity of space partition wield a joint effect on the index performance.

Compact B+-tree [7] is variant of B+-tree which organize data in more compact way via better policy. The basic idea is to use vacant space of the siblings before the overflow happen in the node. Base on this data can accommodate in external structure before splitting operation is require. Figure C and Figure D shows presentation of data sequence {10, 18, 9, 4, 3, 12, 22, 28, 5, 2, 17, 11} for comparison. The result compact B+-tree requires only 6 split and 9 nodes and space utilization is (17/18). On the other hand our conventional B-tree required 9 split and 12 nodes and space utilization is (19/24). This is better policy for the insertion and split operation in traditional index eliminate.

Many other variant of B-tree is also there which are not discussed in this. They are either application specific or data specific.

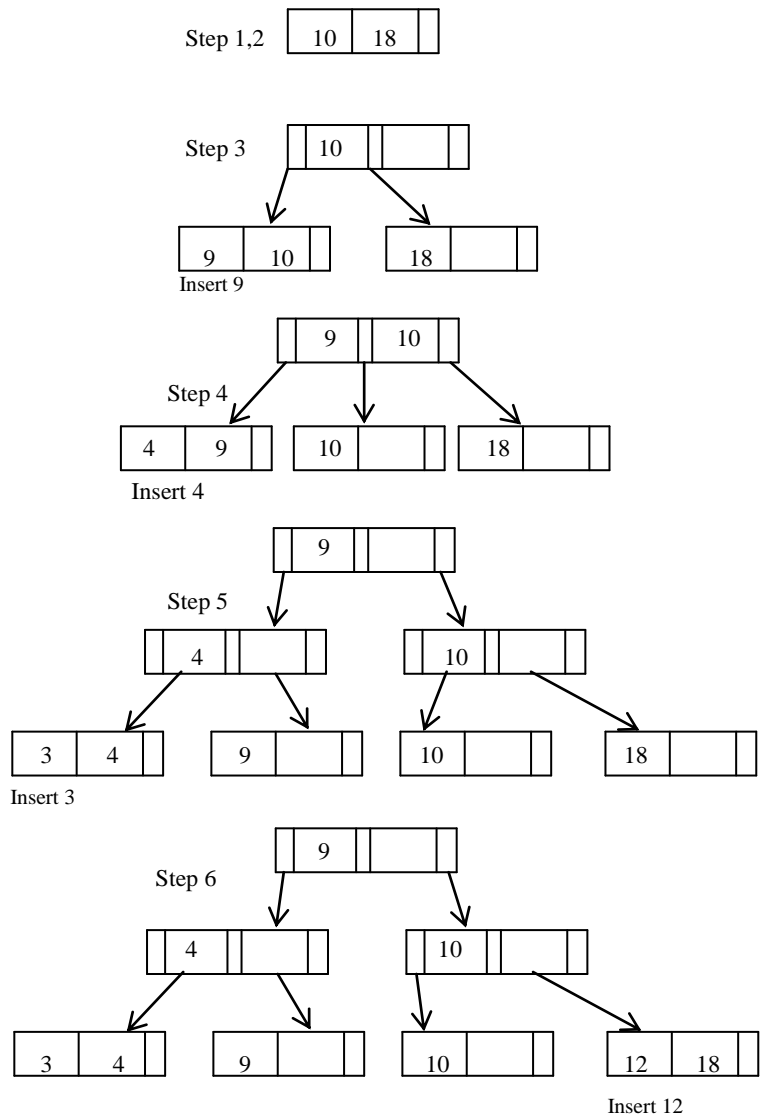

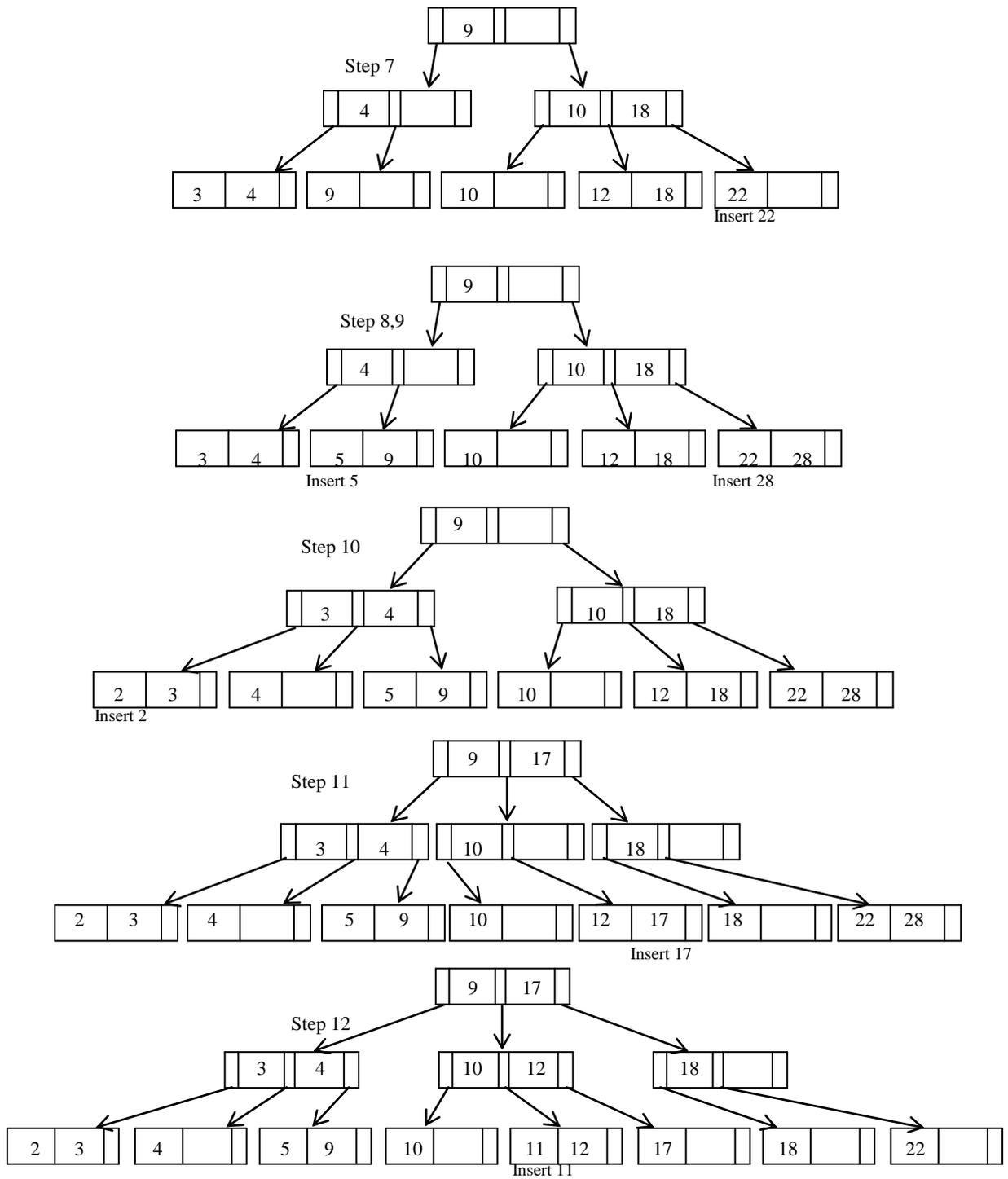

Fig 3: Conventional B+-tree

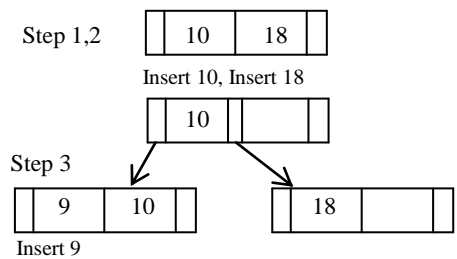

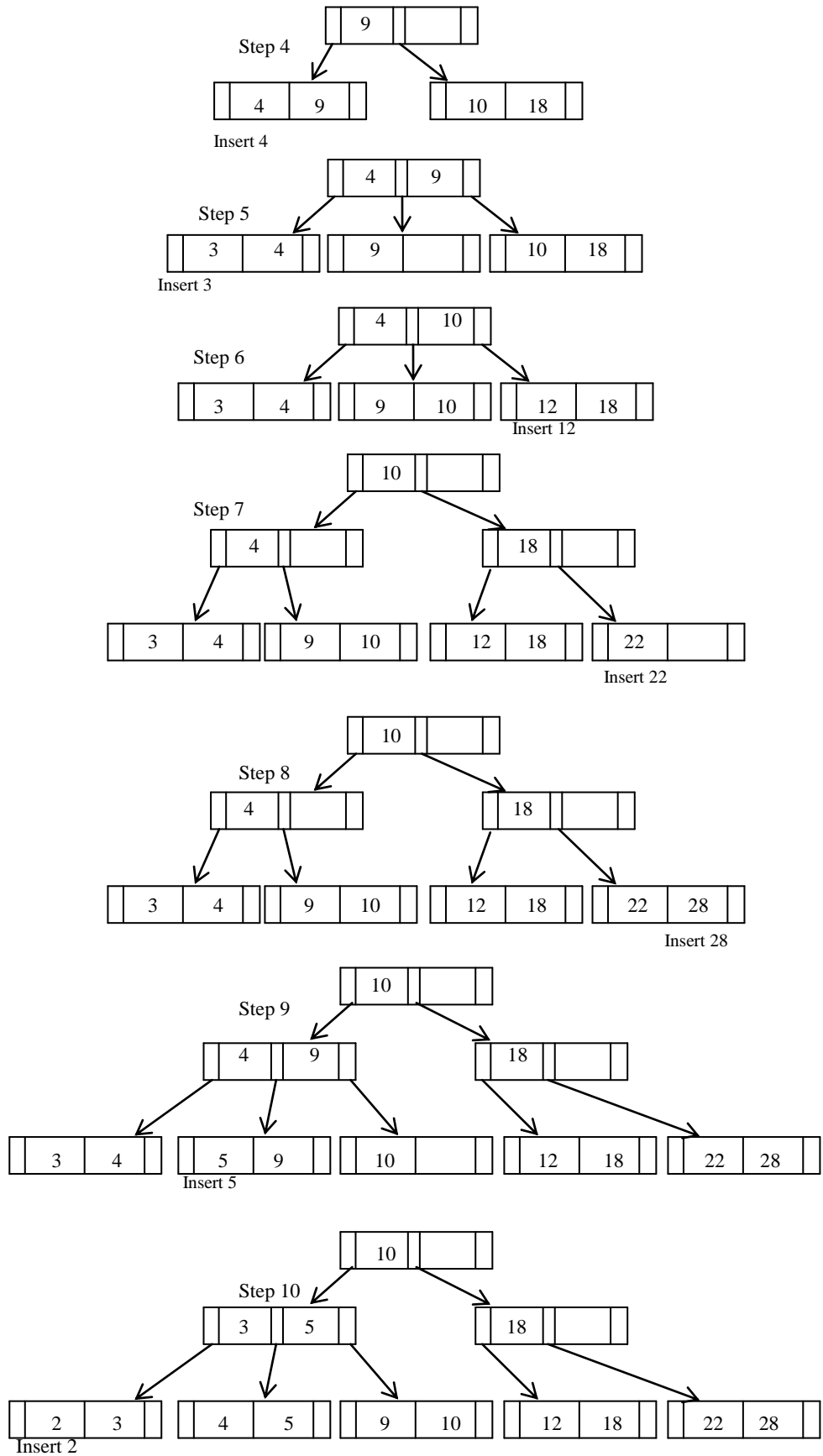

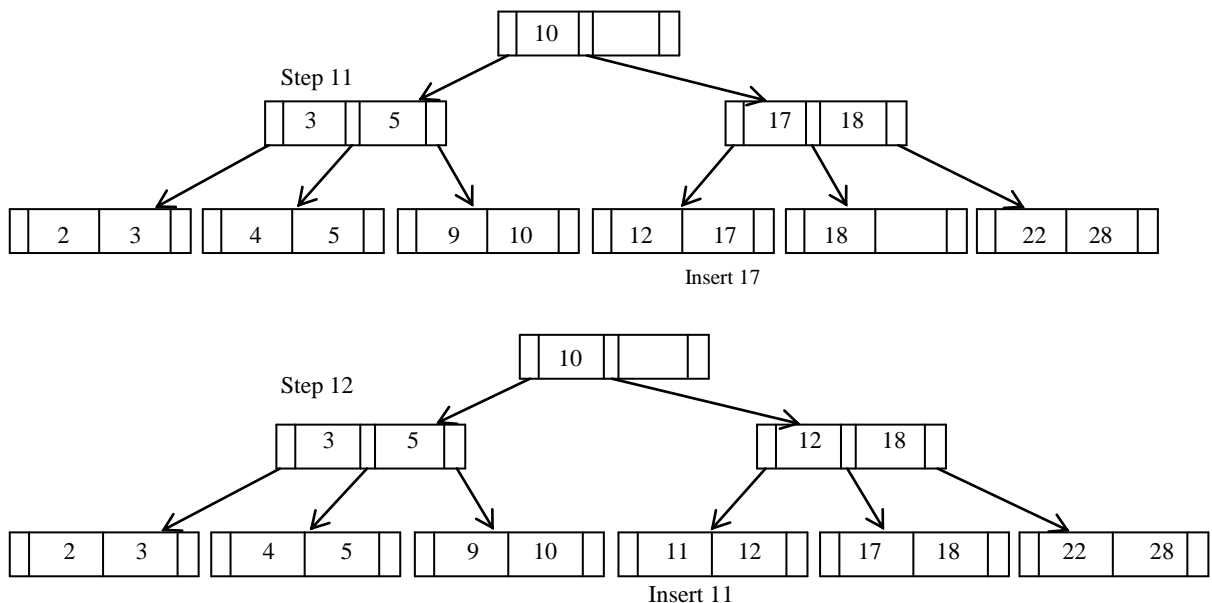

**Fig 4: Compact B+-tree**

## 3.2 Variants of R-tree

R+-tree is a variant of R-tree differs from it in 1. Nodes are not guaranteed to be at least half filled. 2. Entries of internal node do not overlap. 3. Object id may be stored in more than one leaf node. R+-tree searching follows single path fewer nodes are visited than R-tree. But data are duplicated over many leaf node structure of R++-tree can be larger than R-tree. Figure 5 show R+-tree and its relation between rectangles.

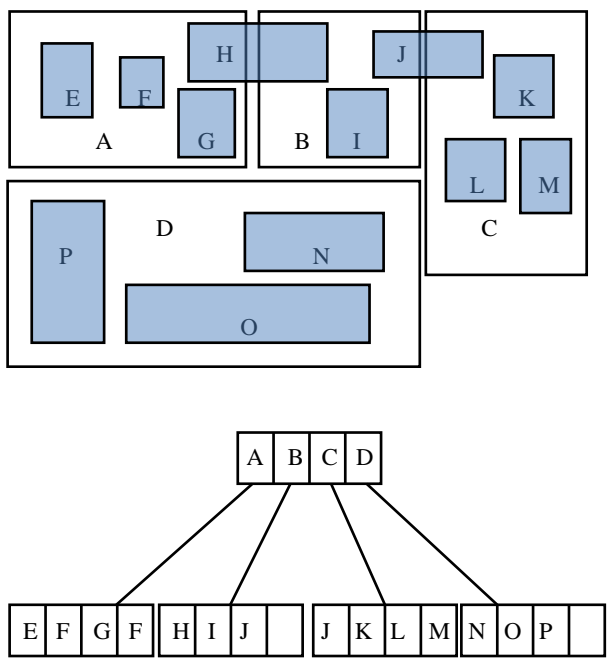

**Fig 5: R+-tree**

R*-tree [11] is also variant of R-tree its results shows that it outperform the traditional R-tree in query processing and performance. It tested parameter area, margin and overlap in different combination. To calculate overlap at each entry and with very distance rectangles probability of overlap is very small. For splitting of node R*-tree first sort lower values and then upper values of the rectangles then two group are determined. Choose goodness of value for the final distribution of entries. Three goodness value and different approaches using them in different combination are tested. 1. Area-value, 2.Margin-value, 3.Overlap-value. R*-tree is very robust in compare to other ugly data distribution. It's one of costly operation is reinsertion but it reduce the split operation. Storage utilization is higher than variants of R-tree but implementation cost is higher than R-tree.

X-tree [14] and M-tree [10] are other variants of R-tree use for the same multidimensional data. Construction of M-tree is fully parametric on distance function d and triangle inequality for efficient queries. It has overlap of regions and no strategy to avoid overlap. Each node there is radios r, every node n and leaf node l residing in node N is at most distance r from N. It is balanced tree and not requires periodical reorganization.

X-tree prevents overlapping of bounding boxes. Which is problem in high dimension, node not split will be result into super-nodes and in some extreme cases tree will linearize.

Hilbert R-tree [5], R-tree variant is used for indexing of object like line, curve, 3-D object and high dimension future based parametric objects. It use quad tree and z-ordering, quad tree divides object into quad tree blocks and increase no of item. It use space filling curves and specifically the Hilbert curve achieve best clustering Figure 6 [5] show Hilbert curve. These goals can achieve for every node (a) store MBR (minimum bounding rectangle), (b) the Largest Hilbert Value of the data rectangles that being to the sub tree with root [5]. Leaf node entries of the form (R, obj_id) where R is MBR of real object and obj_id is pointer to object record. A non- leaf node entries of the form (R, ptr, LHV) where R is MBR, ptr is pointer to child node and LHV is Largest Hilbert value among data rectangle enclose by R. It give 285 of the saving over the best competitor R*-tree on Real data.

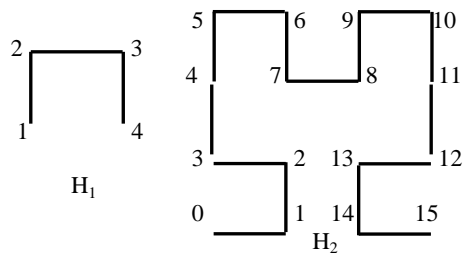

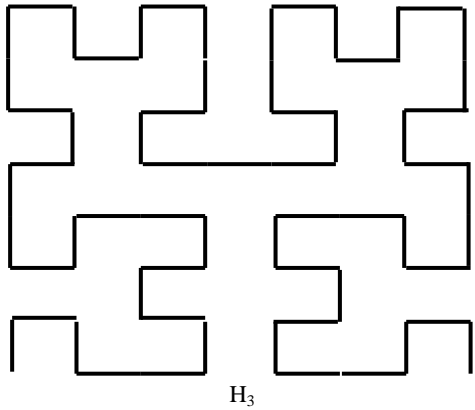

**Fig 6 : Hilbert curves of order 1,2 and 3**

Bloom filter base R-tree (BR-tree) [19] in which bloom filter is integrated to R-tree node. BR-tree is basically R-tree structure for supporting dynamic indexing. In it each node maintains range index to indicate attribute of existing item.

Range query and cover query supported because it store item and range of it together. A Bloom filter is a space-efficient data structure to store an index of an item and can represent a set of items as a bit array using several independent hash functions [16]. Figure 7 [19] show proposed BR-tree structure. BR-tree node is combination of R-tree node and Bloom filter.

BR-tree is also load balanced tree. Overloaded bloom filter produce high false positive probabilities. It reconfigures the multidimensional range using bounding boxes to cover item. BR-tree support Bound query the first index structure to talk about the bound query. Bound query result into range information of multidimensional attribute of a queried item. It is not trivial because BR-tree maintains advantage of Bloom filter and R-tree both. It mixes the queries like bound query and range query after point query result is positive. BR-tree keep consistency between queried data and the attribute bound in an integrated structure so that fast point query and accurate bound query possible. Figure 8 [19] shows example of multiple queries on BR-tree.

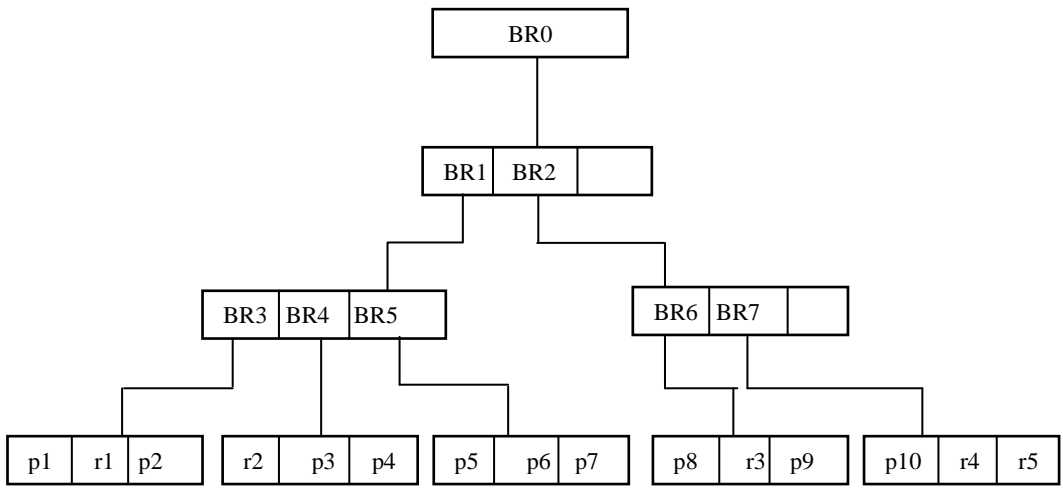

**Fig 7[19] : BR-tree Example**

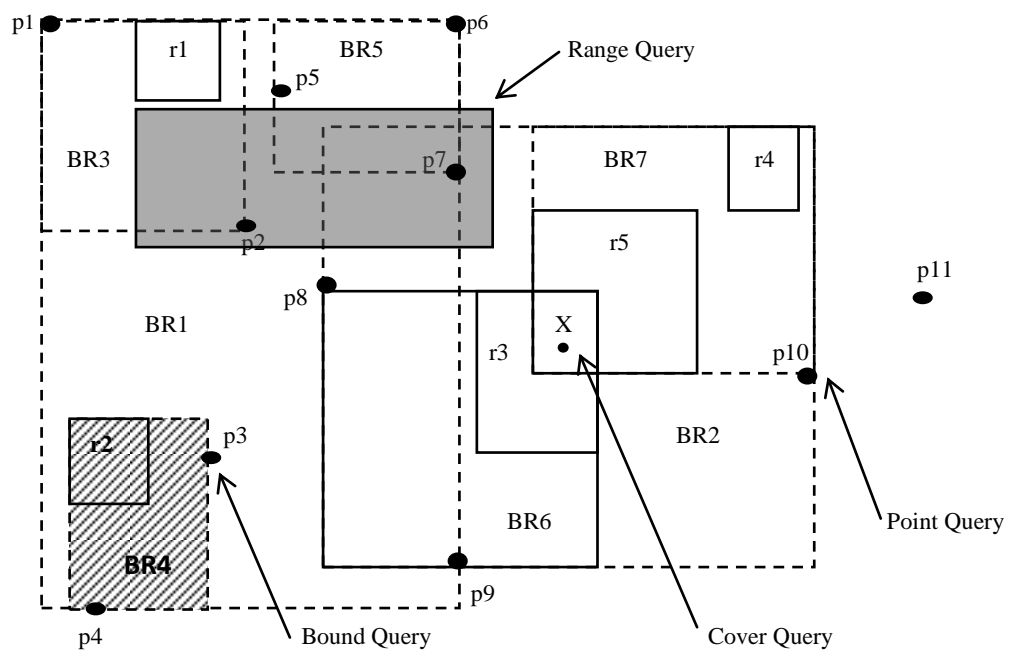

**Fig 8[19] : An example of multiple query in BR-tree**

QR+-tree [7] is hybrid structure of Quad tree (Q-tree) [11] and R-tree. First rough level partition of index space using Q-tree and then use R-tree to index space object. QR+-tree subdivides the spatial area and constructs the first level index. Construction algorithm of second level is improvement splitting algorithm on R-tree. Each quad has a pointer refer to the root and if quad does not have R-tree then pointer will be null. Figure 9 [7] shows the flat chart of QR+-tree and Figure 10 [7] shows the structure chart of QR+-tree. QR+-tree does not have the redundant index information that allows index to store the data directly and save the storage space. Fast and adjustable index makes query processing efficient.

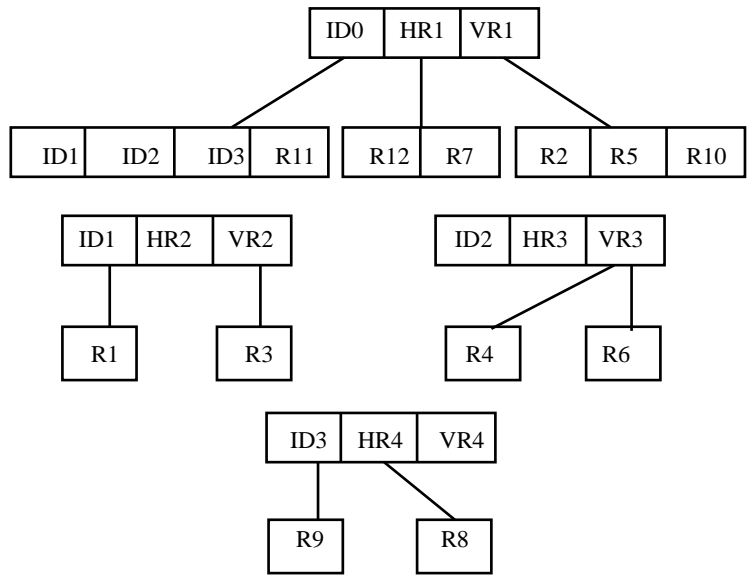

**Fig 9 [7] : Flat chart of QR+-tree**

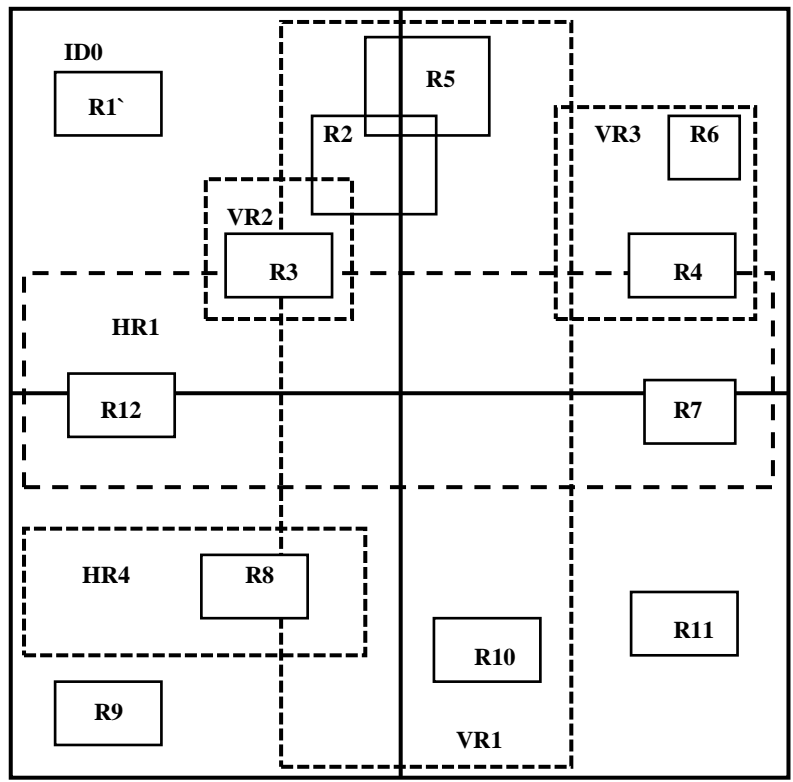

**Fig 10 [7] : Structure chart of QR+-tree**

## 4. COMPARISON BETWEEN INDEX STRUCTURES

### 4.1 Query type

Basically 4 types of query are there Point query, range query, bound query and cover query.

### 4.2 Data type

Two types of data are there linear and multidimensional. Multidimensional data represent the object like curves, rectangles, 3-D objects. Spatial data and high dimensional data are part of multidimensional data.

### 4.3 Complexity

Each and every data structure has complexity in terms of space and time. Most of the index structures have time complexity in terms of O(log n). But different index structures have different factor, terms and condition on algorithm.

### 4.4 Application

Different index structures are used for the different application for the efficient performance and some structures are introduced for the specific application only.

**Table 1. Comparison between Index Structures**

| Index Structure | Query type | Data type | Complexity | Application |
|---|---|---|---|---|
| B-tree | Point query [1] | Linear data [1] | O(log n) | Apple's file system HFS+, Microsoft's NTFS and some Linux file systems, such as btrfs and Ext4. |
| B+-tree | Point query [3] | Linear data [3] | O(log n) | Most of the database management systems like IBM DB2, Microsoft My Sql, Oracle 8, Sybase ASE etc. |
| B*-tree | Point query [3] | Linear data [3] | O(log n) use space more efficiently than B+-tree | HFS and Reiser4 file systems |
| UB-tree | Point query, Range query [18] | Linear data, multidimensional data [18] | O(log n) but not feasible for multidimensional data | Multidimensional range search. |
| H-tree | Point query | Linear data | O(log n) utilize space more efficiently. | Ext3, ext4 Linux file systems. |
| ST$^2$B-tree | Range query, k-NN query [15] | Multidimensional data [15] | Work more efficiently for the moving object data. | Application with multidimensional data but now not use because other data structure outperform it. |
| Compact B-tree | Point query [4] | Linear data [4] | O(log n) but use space more efficiently than B-tree | In place of B-tree. |
| R-tree | Range query [1] | Multidimensional data [1] | Not utilize space more efficiently, not have worst case time complexity. | Real world application like navigation system etc. |
| R+-tree | Range query [16] | Multidimensional data [16] | Non overlapping data utilize space efficiently than R-tree | Multidimensional data object |
| R*-tree | Point query, Range query [9] | Spatial data, multidimensional data [9] | Implementation cost is more than other R-tree variants but robust in data distribution than other ugly structures. | Application with data in form of points and rectangles |
| X-tree | Range query [14] | Multidimensional data, High dimensional data [14] | In some extreme cases tree become linear and time complexity O(n) | High dimension data |
| M-tree | Range query, k-NN query [10] | Multidimensional data [10] | Not require periodic reorganization, time is less in construction. | k-NN query, application use multidimensional (spatial) access methods |

| | | | [10] | |
|---|---|---|---|---|
| Hilbert R-tree | Range queries [5] | Multidimensional data [5] | Search cost give 28% saving above R*-tree. | Cartography, Computer Aided Design(CAD), computer vision and robotics etc. [5] |
| BR-tree | Point query, Range query, Cover query, Bound query [19] | Linear data, multidimensional data [19] | O($\leq$ log n) | Application require all four type of query and also use in distributed environment [19]. |
| QR+-tree | Range query [7] | Large scale spatial data [7] | No redundant information make query processing more efficient. | Large scale GIS database [7]. |

## 5. CONCLUSION AND FUTURE SCOPE

Many variants of B-tree and R-tree are proposed and some of them are used in the real world for the query and performance optimization. Some index structure have less space complexity, some have less time complexity and support different data types. Most of them support point query and single dimensional data efficiently but for range query and multidimensional data specific structure is required and support specific type of data. B-tree and its variants are support point query and single dimensional data efficiently while R-tree and its variants support multidimensional data and range query efficiently. BR-tree support single dimensional, multi-dimensional and all four type of query. New index structure is proposed by making change in previous structure with use of some other data structure like hash function or use two good property of two different structure. Like BR-tree use hash function and QR+-tree use of Q-tree and R-tree. For optimize space complexity change in existing algorithm is made. Like in Compact B-tree. In future take idea from this and change existing index structure. For new index structure change can be made in algorithm, use two different index structure or use data structure or use of data structure like hash in index construction.